\documentclass[letterpaper,10pt]{article}

\usepackage{graphicx}
\usepackage{textcomp}
\usepackage[usenames,dvipsnames]{color}
\usepackage{amsmath}
\usepackage{slashed}
\usepackage[super]{cite}

\def\beq{\begin{equation}}
\def\eeq{\end{equation}}
\def\theta{\vartheta}

\newcommand{\be}{\begin{equation}}
\newcommand{\ee}{\end{equation}}
\newcommand{\ba}{\begin{eqnarray}}
\newcommand{\ea}{\end{eqnarray}}
\newcommand{\lsim}   {\mathrel{\mathop{\kern 0pt \rlap
  {\raise.2ex\hbox{$<$}}}
  \lower.9ex\hbox{\kern-.190em $\sim$}}}
\newcommand{\gsim}   {\mathrel{\mathop{\kern 0pt \rlap
  {\raise.2ex\hbox{$>$}}}
  \lower.9ex\hbox{\kern-.190em $\sim$}}}

\begin{document}

{\center\section*{SEARCH FOR THE FOOTPRINTS OF NEW PHYSICS WITH LABORATORY AND COSMIC NEUTRINOS\footnote{Preprint of a review article solicited for publication in Modern Physics Letters A, submitted for publication.}}
 
\author\tab{{FLOYD W. STECKER}} 
\vspace{12pt}

Astrophysics Science Division, NASA Goddard Space Flight Center, Greenbelt, MD 20771, USA 
\vspace{12pt}

and 
\vspace{12pt}

\center{Department of Physics and Astronomy, UCLA, Los Angeles, CA 90095}}



\begin{abstract}
Observations of high energy neutrinos, both in the laboratory and from cosmic sources, can be a useful probe in searching for new physics. Such observations can provide sensitive tests of Lorentz invariance violation (LIV), which may be a the result of quantum gravity physics (QG).  We review some observationally testable consequences of LIV using effective field theory (EFT) formalism. To do this, one can postulate the existence of additional small LIV terms in free particle Lagrangians, suppressed by powers of the Planck mass. The observational consequences of such terms are then examined. In particular, one can place limits on a class of non-renormalizable, mass dimension five and six Lorentz invariance violating operators that may be the result of QG.  
 

\end{abstract}


\section{Introduction}

General relativity (GR) and quantum field theory (QFT) have been the cornerstones of physics in the 20th and 21st century. GR has provided a deep conceptual framework for understanding such phenomena as black holes, large scale cosmology and gravitational waves, in addition to explaining specific topics such as the orbit of mercury, gravitational redshifts, and gravitational lensing. While being counterintuitive in many respects, QFT has nevertheless provided a deep understanding of physics on the small scale, both accurately describing the interactions of  subatomic particles and as a framework that describes macroscopic emergent phenomena in condensed matter. However, GR and QFT, as they presently stand, are incomplete, being incompatible at the Planck scale of $\lambda_{Pl} = \sqrt{G\hbar/c^3} \sim 10^{-35}$ m~\cite{pl99}, corresponding to energy scale of $1.22 \times 10^{19}$ GeV. In some efforts to unify quantum mechanics with general relativity, many quantum gravity models introduce drastic modifications to space-time at the Planck scale (see, e.g., Ref. 2). Such proposed attempts are theories postulating extra dimensions or a fundamental discreteness of space-time (loop quantum gravity). 
 
One possible modification to space-time structure that has received quite a bit of attention is the idea that Lorentz symmetry is not an exact symmetry of nature. Lorentz symmetry violation has been explored within the context of string theory~\cite{ko89}, loop quantum gravity, Ho\v{r}ava-Lifshitz gravity, causal dynamical triangulations, non-commutative geometry, doubly special relativity, among others (see, e.g., Refs. 4 and 5 and references therein).

While it is not possible to investigate space-time physics at the Planck energy of $\sim 10^{19}$ GeV directly, Planck-scale physics may leave a "footprint" at energies well below the Planck scale, particularly in the form of Lorentz invariance violating (LIV) phenomena. Such lower energy potentially testable effects have been predicted to arise from LIV as traces that originate in physics at the Planck scale. The subject of searching for LIV has generated much interest among both particle physicists and astrophysicists. 

This paper reviews the topic of LIV in the neutrino sector, using both high laboratory neutrino results and astrophysical observations of high energy cosmic neutrinos in order to search for footprints of Planck-scale physics that may be manifested at energies well below the Planck scale. We will discuss here how LIV effects can be searched for using both neutrino oscillation results and the observations of the energy spectrum high energy cosmic neutrinos obtained by the {\it IceCube} collaboration. We base this discussion within the well delineated framework of the standard model extension (SME) formalism of effective field theory (EFT)~\cite{ck98}. 

In all of the treatments discussed in this review, for simplicity we will assume that there are additional LIV terms in the free particle Lagrangians that are rotationally invariant. We thus neglect the much more numerous anisotropic terms that can arise in the SME formalism~\cite{ck98,cm13} but for which there is no present observational evidence. We further assume that the LIV terms are isotropic in the rest system of the cosmic background radiation (CBR), a preferred system picked out by the universe itself. In actuality, we are moving with respect to this system by a velocity $\sim 10^{-3}$ of the speed of light (hereafter taking $c = \hbar = 1$).

In Sect. \ref{form} we discuss the neutrino propagation in the simpler, renormalizable mass dimension 4 model~\cite{co99} ($[d] = 4$). We then discuss the $[d] = 5$ and $[d] = 6$ rotationally invariant operators that are suppressed by one and two factors of the Planck mass respectively. We discuss the effect of such additional terms on neutrino oscillations in 
Sect.~\ref{osc}. We treat electron-positron pair emission by superluminal high energy neutrinos {\it in vacuo} in Sect.~\ref{vpe}. In Sect.~\ref{3n} we treat  neutrino splitting in the $[d] = 5$ and $[d] = 6$ dominant cases. Sect.~\ref{ice} summarizes the recent {\it IceCube} measurements of cosmic high energy neutrinos. Sect.~\ref{cos} places extragalactic neutrino production and propagation in a cosmological framework. In Sect.~\ref{spec} we discuss the effects of LIV on the neutrino spectrum and compare the resulting spectrum with the {\it IceCube} data, placing limits on the strength of the $[d] = 4$ and $[d] = 6$ operators, possibly ruling out dominance of $\cal{CPT}$ violation from a $[d] = 5$ five operator. Sect.~\ref{sum} summarizes the results for superluminal neutrinos. Sect.~\ref{pi} discusses the possibility of stable pions. Sect.~\ref{tel} mentions future observational tests.

\section{Free particle propagation and modified kinematics}
\label{form}

In the effective field theory (EFT) formalism, LIV can be incorporated by the addition of
terms in the free particle Lagrangian that explicitly break Lorentz invariance. 
Since it is well known that Lorentz invariance holds quite well at accelerator
energies, the extra LIV terms in the Lagrangian must be very small. The EFT is considered
an approximation to a true theory that holds up to some limiting high energy (UV) scale.

\subsection{Mass dimension [d] = 4 LIV with rotational symmetry}

For an introduction demonstrating how LIV terms affect particle kinematics,
we consider the simple example of a free scalar particle Lagrangian with an additional small
dimension-4 Lorentz violating term, assuming rotational symmetry~\cite{co99}.  
\begin{equation}
\Delta \mathcal{L}_f = \partial_{i}\Psi^{*}{\bf \epsilon}\partial^{i}\Psi.
\end{equation}
This leads to a modified propagator for a particle of mass $m$
\begin{equation}
-iD^{-1}~=~ (p_{(4)}^2~-~m^2)~+~\epsilon p^2.
\end{equation}
so that we obtain the dispersion relation
\begin{equation}
p_{(4)}^2~=~E^2~-~p^2~ \Rightarrow~ m^2 ~ +~ \epsilon p^2.
\label{LIVdispersion}
\end{equation}

In this example, the low energy "speed of light" maximum attainable particle velocity, here equal to 1 by convention, is replaced by a new maximum attainable velocity (MAV) as $v_{MAV} \neq 1$, which is 
changed by $\delta v \equiv \delta = \epsilon/2.$ 
\begin{equation}
{{\partial  E}\over{\partial |\vec{p}|}}  = {{|\vec{p}|}  \over {\sqrt
{|\vec{p}|^2 + m^2 v_{MAV} ^2}}} v_{MAV},
\label{groupvel}
\end{equation}
\noindent which goes  to $v_{MAV}$ at relativistic energies, $|\vec{p}|^2 \gg m^2$.

For the [d] = 4 case, the superluminal velocity of particle $I$ that is produced by the existence of one or more LIV terms in the free particle Lagrangian will be denoted by
\begin{equation}
v_{I, MAV} \equiv 1 + \delta_{I}
\label{v}
\end{equation}
We are always in the relativistic limit $|\vec{p}|^2 \gg m^2$ for both neutrinos and electrons. Thus, the neutrino or electron velocity is just given by equation~(\ref{v}). 

\subsection{Fermion LIV operators with $[d] > 4$ LIV with rotational symmetry in SME.}
In the cases where $[d] > 4$ Planck-suppressed operators dominate, there will be LIV terms that are proportional to $(E/M_{Pl})^n$, where $n = [d] - 4$, leading to values of $\delta_{I}$ that are energy dependent and are taken to be suppressed by appropriate powers of the Planck mass. 

We again assume rotational invariance in the rest frame of the CBR and consider only the effects of Lorentz violation on freely propagating cosmic neutrinos. Thus, we only need to examine Lorentz violating modifications to the neutrino kinetic terms. Majorana neutrino couplings are ruled out in SME in the case of rotational symmetry~\cite{ko12}. Therefore we only consider Dirac neutrinos. 

There are many ways that LIV terms in the free particle Lagrangian can affect neutrino physics.  We will here consider only two of these consequences, viz., (1) their effect on modifying neutrino oscillations and (2) the effect of resulting changes in the kinematics of particle interactions. These changes can modify the threshold energies for particle interactios, allowing or forbidding such interactions~\cite{co99,sg01}.  

Using equation (\ref{LIVdispersion}), one can define an effective mass, $\tilde{m}(E)$, that is a useful parameter for analyzing LIV-modified kinematics. The effective mass is constructed to include the effect of the LIV terms. We define an effective mass $\tilde{m}_I(E)$ for a particle for type $I$ using the dispersion relation (\ref{LIVdispersion}) as~\cite{co99,st15}  
\begin{equation} 
\tilde{m}_{I}^2(E)=m_{I}^2+ 2\delta_I E_{I}^2,
\label{effectivemass}
\end{equation}
where the velocity parameters $\delta_I$ are now energy dependent dimensionless coefficients for each species, $I$, that are contained in the Lagrangian. Also, we define the parameter $\delta_{IJ} \equiv \delta_{I} - \delta_{J}$ as the Lorentz violating difference between the MAVs of particles $I$ and $J$. In general $\delta_{IJ}$ will therefore be of the form
\begin{equation}
\delta_{IJ}=\sum_{n=0,1,2}\kappa_{IJ,n} \left(\frac{E}{M_{Pl}}\right)^n.
\label{d}
\end{equation}

If we wish to assume the dominance of Planck-suppressed terms in the Lagrangian as tracers of Planck scale physics, it follows that that $\kappa_{\nu e,0} \ll \kappa_{\nu e,1}, \kappa_{\nu e,2}$\footnote{Several mechanisms have been proposed for the suppression of the LIV $[d] = 4$ term in the Lagrangian. See, e.g., the review in Ref.~\cite{li13}.} 
Alternatively, we may {\it postulate} the existence of only Planck-suppressed terms in the Lagrangian, i.e., $\kappa_{\nu e,0} = 0$. We can further simplify by noting the important connection between LIV and $\cal{CPT}$ violation. Whereas a local interacting theory that violates $\cal{CPT}$ invariance will also violate Lorentz invariance~\cite{gr02}, the converse does not follow; an interacting theory that violates Lorentz invariance may, or may not, violate $\cal{CPT}$ invariance. LIV terms of odd mass dimension $[d]= 4 + n$ are $\cal{CPT}$-odd and violate $\cal{CPT}$, whereas terms of even mass dimension are $\cal{CPT}$-even and do not violate $\cal{CPT}$~\cite{ko09}. We can then specify a dominant term for $\delta_{IJ}$ in equation (\ref{d}) depending on our choice of $\cal{CPT}$. Considering Planck-mass suppression, the dominant term that admits $\cal{CPT}$ violation is the $n = 1$ term in equation (\ref{d}). On the other hand, if we require $\cal{CPT}$ conservation, the
$n = 2$ term in equation (\ref{d}) is the dominant term. Thus, we can choose as a good approximation to equation (\ref{d}), a single dominant term with one particular power of $n$ by specifying whether we are considering $\cal{CPT}$ even or odd LIV.  As a result, $\delta_{IJ}$ reduces to
\begin{equation}
\delta_{IJ}~ \equiv \kappa_{IJ,n} \left({{E}\over{{M_{Pl}}}}\right)^{n} 
\label{sub}
\end{equation}
with $n = 1$ or $n = 2$ depending on the status of $\cal{CPT}$. We note that in the SME formalism, since odd-[d] LIV operators are $\cal{CPT}$ odd, the $\cal{CPT}$-conjugation property implies that neutrinos can be superluminal while antineutrinos are subluminal or vice versa~\cite{ko12}. This will have consequences in interpreting our results, as we will discuss later.

In equations (\ref{d}) and (\ref{sub}) we have not designated a helicity index on the $\kappa$ coefficients. The fundamental parameters in the Lagrangian are generally helicity dependent~\cite{ja03}. In the $n = 1$ case a helicity dependence must be generated in the electron sector due to the $\cal{CPT}$ odd nature of the LIV term. However, the constraints on the electron coefficient are extremely tight from observations of the Crab nebula~\cite{st14b}. Thus, the contribution to $\kappa_{\nu e,1}$ from the electron sector can be neglected. In the $n = 2$ case, which is $\cal{CPT}$ even,we can set the left and right handed electron coefficients to be equal by imposing parity symmetry~\cite{st15}.
 
\section{LIV in the neutrino sector I - Neutrino Oscillations}
\label{osc}

We now consider the effect on neutrino oscillations of Lorentz violating terms in the Lagrangian that are suppressed by powers of the Planck mass. We again note that the dominant term that admits $\cal{CPT}$ violation is the $n = 1$ term in equation (\ref{d}); the dominant term that conserves $\cal{CPT}$ is the $n = 2$ term in equation (\ref{d}). Given Planck mass suppression, we choose one of these two terms to be the single dominant term with one particular power of $n$, depending on whether $\cal{CPT}$ is conserved or not.  As a result, $\delta_{IJ}$ is given by equation (\ref{sub}). We take the effective mass, $\tilde{m}$, to be given by equation (\ref{effectivemass}) where the $m_i$ denotes one of the three possible mass eigenstates of the neutrino.

We then consider a neutrino with flavor $I$ with  momentum $p$ transitioning into flavor $J$. The amplitude for this neutrino to be in a mass eigenstate $i$ is then denoted by the matrix $U_{Ii}$ where ${\bf U}$ denotes the unitary matrix, $\sum U^{\dag}_{Ji}U_{Ii} = \delta_{IJ}$, with $\delta$ here denoting the delta function as opposed to the definition given in the previous section. 

These considerations change the usual relations for neutrino oscillations. For example, in the case of atmospheric $\nu_{\mu}$ oscillations, the survival probability in the Lorentz invariant case is given by 

\begin{equation}
P_{\nu_{\mu}} \simeq 1 - sin^2(2\theta_{23})sin^2\left({\Delta m^2_{atm}L}\over{4E}\right)
\label{atmLI}
\end{equation}

We can include the effect of an LIV term in the Lagrangian by making the substitution
\begin{equation}
m^2_{atm} \rightarrow \tilde{m}^2_{atm}(E) =m^2_{atm} + 2\delta_{ij} E^2
\end{equation}

\noindent where $\tilde{m}^2(E)$ is given by equation (\ref{effectivemass}). In that case, we find that equation (\ref{atmLI}) is modified by an additional LIV term proportional to the difference in neutrino velocities, $\delta_{IJ}$. It immediately follows that
\begin{equation}
P_{\nu_{\mu}} \simeq 1 - sin^2(2\theta_{23})sin^2\left({{\Delta m^2_{atm}L}\over{4E}} + 
{{\delta_{IJ}}}{{EL\over{2}}}\right)
\label{atmLIV}
\end{equation}

\noindent where the square of the difference between the mass eigenvalues $\Delta m^2_{atm} = \Delta m^2_{31}$~\cite{co99,di12,mac13}. Recent results on atmospheric neutrino oscillations \cite{gg04,ab15,gg16} (also Gonzalez-Garcia and Maltoni, private communication) give an upper limit for the difference in velocities between the $\nu_{\mu}$ and $\nu_{\tau}$ neutrinos, $\delta_{\nu_{\mu} \nu_{\tau}} < \cal{O}$ $(10^{-26})$. It is interesting to note that if Lorentz invariance is violated, equation (\ref{atmLIV}) implies that neutrino oscillations would occur even if neutrinos are massless or if the square of their mass differences is zero. The LIV term in equation (\ref{atmLIV}) can dominate at very high energies and very large distances~\cite{di12} , as is the case for many astrophysical applications. This dominance can be even more profound for high mass dimensions where $\delta_{IJ}$ is given by equation (\ref{sub}).

\section{LIV in the neutrino sector II - Lepton Pair Emission}
\label{vpe}

Since the Lorentz violating operators change the free field behavior and dispersion relation, interactions such as fermion-antifermion pair emission by slightly superluminal neutrinos become kinematically allowed~\cite{co99,co11} and can thus cause significant observational effects. An example of such an interaction is $\nu_e$ "splitting", i.e., $\nu_e \rightarrow \nu_e + \nu_i + \bar{\nu_i}$ where $i$ is a flavor index.  Neutrino splitting can be represented as a rotation of the Feynman diagram for neutrino-neutrino scattering  which is allowed by relativity.  However, absent a violation of Lorentz invariance, neutrino splitting is forbidden by conservation of energy and momentum.  In the the case of LIV allowed superluminal neutrinos the dominant pair emission reactions are neutrino splitting and its close cousin, vacuum electron-positron pair emission (VPE) $\nu_i \rightarrow \nu_i + e^+ + e^-$, as these are the reactions with the lightest final state masses. We now set up a simplified formalism to calculate the possible observational effect of these two specific anomalous interactions on the interpretation of the neutrino spectrum observed by the {\it IceCube}
collaboration.

\subsection{Lepton pair emission in the [d] = 4 case}

\begin{figure}[t!]
\includegraphics[width=4.8in]{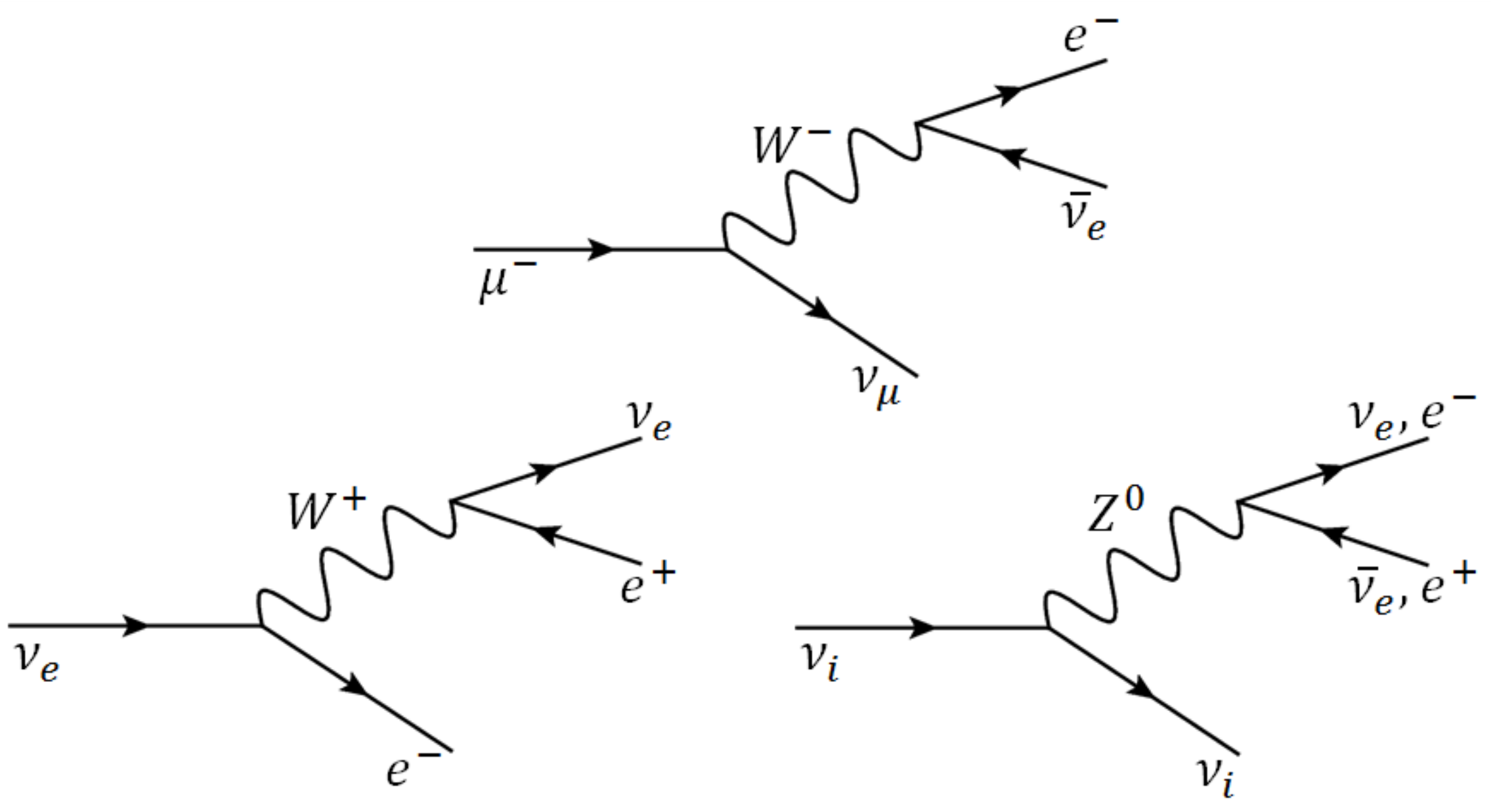}
\caption{Diagrams for muon decay (top), charged current mediated VPE (bottom left), and neutral current mediated neutrino splitting/VPE (bottom right).  Time runs from left to right and the flavor index $i$ represents $e,\mu$, or $\tau$ neutrinos.}\label{fig:diagrams}
\end{figure}

In this section we consider the constraints on the LIV parameter $\delta_{\nu e}$. We first relate the rates for superluminal neutrinos with that of a more familiar tree level, weak force mediated standard model decay process: muon decay, $\mu^-\rightarrow \nu_\mu + \bar{\nu}_e + e^-$, as the process are very similar (see Figure~\ref{fig:diagrams}).  

For muons with a Lorentz factor $\gamma_{\mu}$ in the observer's frame the decay rate is found to be
\begin{equation}
{ \Gamma \  =   \gamma_{\mu}^{-1}} {{G_F^2 m_{\mu}^5}\over{192\pi^3}}
\label{mu}
\end{equation}
where $G_F^2 = g^4/(32M_{W}^4)$, is the square of the Fermi constant equal to $1.360 \times 10^{-10} \ {\rm GeV}^{-4}$, with $g$ being the weak coupling constant and $M_{W}$ being the $W$-boson mass in electroweak theory.

We apply the effective energy-dependent mass-squared formalism given by equation (\ref{effectivemass}) to determine the scaling of the emission rate with the $\delta$ parameter and with energy. Noting that for any reasonable neutrino mass, $m_{\nu} \ll 2\delta_{\nu e} E_{\nu}^2$, it follows that $\tilde{m_{\nu}}^2(E) \simeq 2\delta_{\nu e} E_{\nu}^2$.\footnote{At relativistic energies, assuming that the Lorentz violating terms yield small corrections to $E$ and $p$, it follows that $E \simeq p$.}
 
We therefore make the substitution
\begin{equation}
m_{\mu}^2 \ \rightarrow \ \tilde{m_{\nu}}^2(E) \simeq 2\delta_{\nu e} E_{\nu}^2 
\end{equation}
\noindent from which it follows that
\begin{equation}
{\gamma_{\mu}^2} \ \rightarrow \ {{E_{\nu}^2}\over{2\delta_{\nu e} E_{\nu}^2}} = (2\delta_{\nu e})^{-1}.
\end{equation}
The rate for the vacuum pair emission processes (VPE) is then
\begin{equation}
\Gamma \ \propto \ (2\delta_{\nu e})^{1/2} G_F^2 (2\delta_{\nu e} E_{\nu}^2)^{5/2}  
\label{dimension}
\end{equation}
\noindent which gives the proportionality
\begin{equation}
\Gamma \ \propto \ G_F^2 \ \delta_{\nu e}^3 E_{\nu}^5 
\label{prop}
\end{equation}
\noindent showing the strong dependence of the decay rate on both $\delta_{\nu e}$ and $E_{\nu}$.

The energy threshold for $e^+e^-$ pair production is given by\cite{sg01}
\begin{equation}
E_{th} = m_e\sqrt{{{2}\over{{\delta_{\nu e}}}}} 
\label{threshold}
\end{equation}
with $\delta \equiv \delta_{\nu e}$ given by equation (\ref{sub}) the rate for the VPE process, $\nu \to \nu \,e^+\, e^-$ via the neutral current $Z$-exchange channel, has been calculated to be~\cite{co11} 
\begin{equation}
\Gamma = \frac{1}{14}\frac{G_F^2 (2\delta)^3E_{\nu}^5}{192\,\pi^3} = 1.31 \times 10^{-14} \delta^3 E_{\rm GeV}^5\ \ {\rm GeV}.
\label{G}
\end{equation}
\noindent with the mean fractional energy loss per interaction from VPE of 78\%~\cite{co11}. 

In general, the charged current $W$-exchange channels contribute as well.  However, this channel is only kinematically relevant for $\nu_e$'s, as the production of $\mu$ or $\tau$ leptons by $\nu_{\mu}$'s or $\nu_{\tau}$'s has a much higher energy threshold due to the larger final state particle masses (equation (\ref{threshold}) with $m_e$ replaced by $m_{\mu}$ or 
$m_{\tau}$), with the neutrino energy loss from VPE being highly threshold dependent. Owing to neutrino oscillations, neutrinos propagating over large distances spend 1/3 of their time in each flavor state. Thus, the flavor population of neutrinos from astrophysical sources is expected to be [$\nu_e$:$\nu_{\mu}$:$\nu_{\tau}$] = [1:1:1] so that CC interactions involving $\nu_e$'s will only be important 1/3 of the time. 

The vacuum \v{C}erenkov emission (VCE) process, $\nu \rightarrow \nu + \gamma$, is also kinematically allowed for superluminal neutrinos. However, since the neutrino has no charge, this process entails the neutral current channel production of a loop consisting of a virtual electron-positron pair followed by its annihilation into a photon. Thus, the rate for VCE is a factor of $\alpha$ lower than that for VPE. 

Neutrino pair emission, a.k.a. neutrino splitting, is unimportant for energy loss of superluminal neutrinos in the $[d] = 4$ case because the fractional energy loss per interaction is very low~\cite{co11} owing to the small
velocity difference between neutrino flavors obtained from neutrino oscillation data (See Section \ref{osc}). However, this is not true in the cases with $[d] > 4$.

\subsection{Vacuum $e^+e^-$ Pair Emission in the $[d] > 4$ cases}.

Using equations (\ref{sub}) and (\ref{prop}) and the dynamical matrix element taken from the simplest case~\cite{ca12}, we can generalize equation (\ref{G}) for arbitrary values of $n = [d] - 4$~\cite{st15}.
\begin{equation}
\Gamma = \frac{G_F^2}{192\,\pi^3}[(1-2s_W^2)^2 + (2s_W^2)^2]\zeta_n \kappa_n^3 \frac{E_\nu^{3n+5}}{M_{Pl}^{3n}} 
\label{G2}
\end{equation} 
\noindent where $s_W$ is the sine of the Weinberg angle ($s_W^2 = 0.231$) and the $\zeta_n$'s are  numbers of order $1$~\cite{ca12}.
 
For the $n = 1$ case we obtain the VPE rate
\begin{equation}
\Gamma = 1.72 \times 10^{-14} \kappa_1^3 E_{\rm GeV}^5\ (E/M_{Pl})^3 \ {\rm GeV},
\label{Gn1}
\end{equation}
and for the $n = 2$ case we obtain the VPE rate
\begin{equation}
\Gamma = 1.91 \times 10^{-14} \kappa_2^3 E_{\rm GeV}^5\ (E/M_{Pl})^6 \ {\rm GeV}.
\label{Gn2}
\end{equation}
 
\section{LIV in the neutrino sector III - decay by neutrino pair emission (neutrino splitting)}
\label{3n} 

The process of neutrino splitting in the case of superluminal neutrinos, i.e., $\nu \rightarrow 3\nu$ is relatively unimportant in the $[d] = 4, n = 0$ owing to the small
velocity difference between neutrino flavors obtained from neutrino oscillation data (see Section \ref{osc}). However, this is not true in the cases with $[d] > 4$. In the presence of $[d] > 4$ $(n > 0)$ terms in a Planck-mass suppressed EFT, the velocity differences between the neutrinos, being energy dependent, become significant~\cite{mac13}. The daughter neutrinos travel with a smaller velocity. The velocity dependent energy of the parent neutrino is therefore greater than that of the daughter neutrinos.  Thus, the neutrino splitting becomes kinematically allowed. Let us then consider the $n = 1$ and $n = 2$ scenarios. 

Neutrino splitting is a neutral current (NC) interaction that can occur for all 3 neutrino flavors. The total neutrino splitting
rate obtained is therefore three times that of the NC mediated VPE process above threshold. Assuming the three daughter neutrinos
each carry off approximately 1/3 of the energy of the incoming neutrino, then 
for the $n = 1$ case one obtains the neutrino splitting rate~\cite{st15}
\begin{equation}
\Gamma = 5.16 \times 10^{-14} \kappa_1^3 E_{\rm GeV}^5\ (E/M_{Pl})^3 \ {\rm GeV},
\label{Gsplitn1}
\end{equation}
and for the $n = 2$ case we obtain the neutrino splitting rate
\begin{equation}
\Gamma = 5.73 \times 10^{-14} \kappa_2^3 E_{\rm GeV}^5\ (E/M_{Pl})^6 \ {\rm GeV}.
\label{Gsplitn2}
\end{equation}

The threshold energy for neutrino splitting is proportional to the neutrino mass so that it is negligible compared to that given by equation ({\ref{threshold}).

\section{The neutrinos observed by {\it IceCube}} 
\label{ice}

As of this writing, the {\it IceCube} collaboration has identified $87_{-10}^{+14}$ events from neutrinos of astrophysical origin with energies above 10 TeV, with the error in the number of astrophysical events determined by the modeled subtraction of both conventional and prompt atmospheric neutrinos and also penetrating atmospheric muons at energies below 60 TeV~\cite{aa15}. 

There are are four indications that the the bulk of cosmic neutrinos observed by {\it IceCube} with energies above 0.1 PeV are of extragalactic origin: (1) The arrival distribution of the reported events with $E > 0.1$ PeV observed by {\it IceCube} above atmospheric background is consistent with isotropy~\cite{aa13,aa14,aa15}.
(2) At least one of the ~PeV neutrinos came from a direction off the galactic plane~\cite{aa15}.
(3) The diffuse galactic neutrino flux is expected to be well below that observed by {\it IceCube}~\cite{st79}. (4) Upper limits on diffuse galactic $\gamma$-rays in the TeV-PeV energy range imply that galactic neutrinos cannot account for the neutrino flux observed by {\it IceCube}~\cite{ah14}.

Above 60 TeV, the {\it IceCube} data are roughly consistent with a spectrum given by $E_{\nu}^2(dN_{\nu}/dE_{\nu}) \simeq \ 10^{-8} \ {\rm GeV}{\rm cm}^{-2}{\rm s}^{-1}$~\cite{aa13,aa14,aa15}. However, {\it IceCube} has not detected any neutrino induced events from the Glashow resonance effect at 6.3 PeV. In this effect, electrons in the {\it IceCube} volume provide enhanced target cross sections for $\bar{\nu}_{e}$'s through the $W^-$ resonance channel, $\bar{\nu}_{e} + e^- \rightarrow W^- \rightarrow shower$, at the resonance energy $E_{\bar{\nu}_{e}} = M_W^2/2m_{e} = 6.3$ PeV~\cite{gl60}. The enhancement from the Glashow resonance effect is expected to be about a factor of $\sim 10$~\cite{aa13}. Owing to oscillations it is expected that 1/3 of the potential 6.3 PeV neutrinos would be ${\nu}_{e}$'s plus $\bar{\nu}_{e}$'s unless new physics is involved.

Thus, the enhancement in the overall effective area expected is a factor of $\sim$3. Taking account of the increased effective area between 2 and 6 PeV and a decrease from an assumed neutrino energy spectrum of $E_{\nu}^{-2}$, we would expect about 3 events at the Glashow resonance energy, provided that the number of $\bar{\nu}_{e}$'s is equal to the number of ${\nu}_{e}$'s. Even without considering the resonance effect, several neutrino events above 2 PeV would be expected if the $E_{\nu}^{-2}$ spectrum extended to higher energies. Thus, the lack of this flux of neutrinos above $\sim 2$ PeV energy and at the 6.3 PeV resonance may be indications of a cutoff in the neutrino spectrum. 

\section{Extragalactic Superluminal Neutrino Propagation}
\label{cos}

Monte Carlo techniques have been employed to determine the effect of neutrino splitting and VPE on putative superluminal neutrinos~\cite{st15,st14}. Extragalactic neutrinos were propagated from cosmological distances taking account of the resulting energy loss effects by VPE and redshifting in the $[d] = 4$ case. In the $[d] > 4$ cases energy loses include those from both neutrino splitting and VPE, as well as redshifting. It was assumed that the neutrino sources have a redshift distribution similar to that of the star formation rate~\cite{be13}. Such a redshift distribution appears to be roughly applicable for both active galactic nuclei and $\gamma$-ray bursts. A simple neutrino source spectrum proportional to $\sim$ $E^{-2}$ was assumed between 100 TeV and 100 PeV, as is the case for cosmic neutrinos observed by {\it IceCube} with energies above 60 TeV~\cite{aa14}. The final results on the propagated spectrum were normalized to an energy flux of $E_{\nu}^2(dN_{\nu}/dE_{\nu}) \simeq 10^{-8}  {\rm GeV}{\rm cm}^{-2}{\rm s}^{-1}{\rm sr}^{-1}$, consistent with the {\it IceCube} data for both the southern and northern hemisphere. Results were obtained for VPE threshold energies between 10 PeV and 40 PeV as given by equation ({\ref{threshold}), corresponding to values of $\delta_{\nu e}$ between $5.2\times10^{-21} \ {\rm and}~ 3.3 \times 10^{-22}$. 

Since the neutrinos are extragalactic and survive propagation from all redshifts, cosmological effects must be taken into account in deriving new LIV constraints. Most of the cosmic PeV neutrinos will come from sources at redshifts between $\sim$0.5 and $\sim$2~\cite{be13}. The effect of the cosmological $\Lambda$CDM redshift-distance relation is given by
\beq
D(z) = {{c}\over{H_0}}\int\limits_0^z\frac{dz'}{(1+z')\sqrt{\Omega_{\rm\Lambda} + \Omega_{\rm M} (1 + z')^3}}
\eeq 
\noindent where the Hubble constant $H_0 =$ 67.8 km s$^{-1}$ Mpc$^{-1}$, $\Omega_{\rm\Lambda}$ = 0.7, and $\Omega_{\rm M}$ = 0.3. 

The energy loss due to redshifting is given by
\begin{equation}
-(\partial \log  E/\partial t)_{redshift} =  H_{0}\sqrt{\Omega_{m}(1+z)^3 +
  \Omega_{\Lambda}}.
\label{redshift}  
\end{equation}
The decay widths for the VPE process are given by equations (\ref{Gn1}) and (\ref{Gn2}) for the cases $n = 1$ and $n = 2$ respectively while those for neutrino splitting are given by equations (\ref{Gsplitn1}) and (\ref{Gsplitn2}). 

\section{The theoretical neutrino energy spectrum}
\label{spec}

\subsection{[d] = 4 $\cal{CPT}$ Conserving Operator Dominance}

In their seminal paper, using equation (\ref{G}), Cohen and Glashow~\cite{co11} showed how the VPE process in the $[d] = 4$ case implied powerful constraints on LIV. They obtained an upper limit of $\delta < \cal{O}$ $(10^{-11})$ based on the initial observation of high energy neutrinos by {\it IceCube}~\cite{ab11}. Further predictions of limits on $\delta$ with cosmological factors taken into account were then made, with the predicted spectra showing a pileup followed by a cutoff~\cite{go12}. An upper limit of $\delta < \cal{O}$ $(10^{-18})$ was obtained~\cite{bo13} based on later {\it IceCube} observations~\cite{aa13}. 

Using the energy loss rate given by equation (\ref{G}), a value for $\delta_{\nu e} < 10^{-20}$ was obtained based on a model of redshift evolution of neutrino sources and using Monte Carlo techniques to take account of propagation effects as discussed in Section \ref{cos}~\cite{st14,st15}. The upper limit on $\delta_{e}$ is given by $\delta_{e} \le 5 \times 10^{-21}$~\cite{st14b}. Taking this into account, one gets the constraint $\delta_{\nu} \le (0.5 - 1) \times 10^{-20}$. The spectra derived therein for the $[d] = 4$ case also showed a pileup followed by a cutoff. The predicted a cutoff is determined by redshifting the threshold energy effect. 

\subsection{[d] = 6 $\cal{CPT}$ Conserving Operator Dominance} 

In both the $[d] = 4$ and $[d] = 6$ cases, the best fit matching the theoretical propagated neutrino spectrum, normalized to an energy flux of $E_{\nu}^2(dN_{\nu}/dE_{\nu}) \simeq 10^{-8}  {\rm GeV}{\rm cm}^{-2}{\rm s}^{-1}{\rm sr}^{-1}$ below 0.3 PeV, with the {\it IceCube} data corresponds to a VPE rest-frame threshold energy $E_{\nu, \rm th} = 10$ PeV, as shown in Figure~\ref{combined}~\cite{st14,st15}. This corresponds to $\delta_{\nu e} \equiv \delta_{\nu} - \delta_e \le \ 5.2 \times 10^{-21}$. Given that $\delta_e \le \ 5 \times 10^{-21}$, it is again found that $\delta_{\nu} \le (0.5 - 1) \times 10^{-20}$. As shown in Figure \ref{thresholdeffects}, values of $E_{\nu, \rm th}$ less than 10 PeV are inconsistent with the {\it IceCube} data. The result for a 10 PeV rest-frame threshold energy is just consistent with the {\it IceCube} results, giving a cutoff effect above 2 PeV.  

In the case of the $\cal{CPT}$ conserving $[d] = 6$ operator (n = 2) dominance, as in the $[d] = 4$ case, the results shown in Figure~\ref{combined} show a high-energy drop off in the propagated neutrino spectrum near the redshifted VPE threshold energy and a pileup in the spectrum below that energy. This predicted drop off may be a possible explanation for the lack of observed neutrinos above 2 PeV~\cite{st14,st15}. This pileup is caused by the propagation of the higher energy neutrinos in energy space down to energies within a factor of $\sim$ 5 below the VPE threshold. 

The pileup effect caused by the neutrino splitting process is more pronounced than that caused by the VPE process because neutrino splitting produces two new lower energy neutrinos per interaction. This would be a potential way of distinguishing a dominance of $[d] > 4$ Planck-mass suppressed interactions from $[d] = 4$ interactions. Thus, with better statistics in the energy range above 100 TeV,a significant pileup effect would be a signal of Planck-scale physics. Pileup features are indicative of the fact that fractional energy loss from the last allowed neutrino decay before the VPE process ceases is 78\%~\cite{co11} and that for neutrino splitting is taken to be 1/3. The pileup effect is similar to that of energy propagation for ultrahigh energy protons near the GZK threshold~\cite{st89}.

\begin{figure}[!t]
\centerline{\includegraphics[width=4.5in]{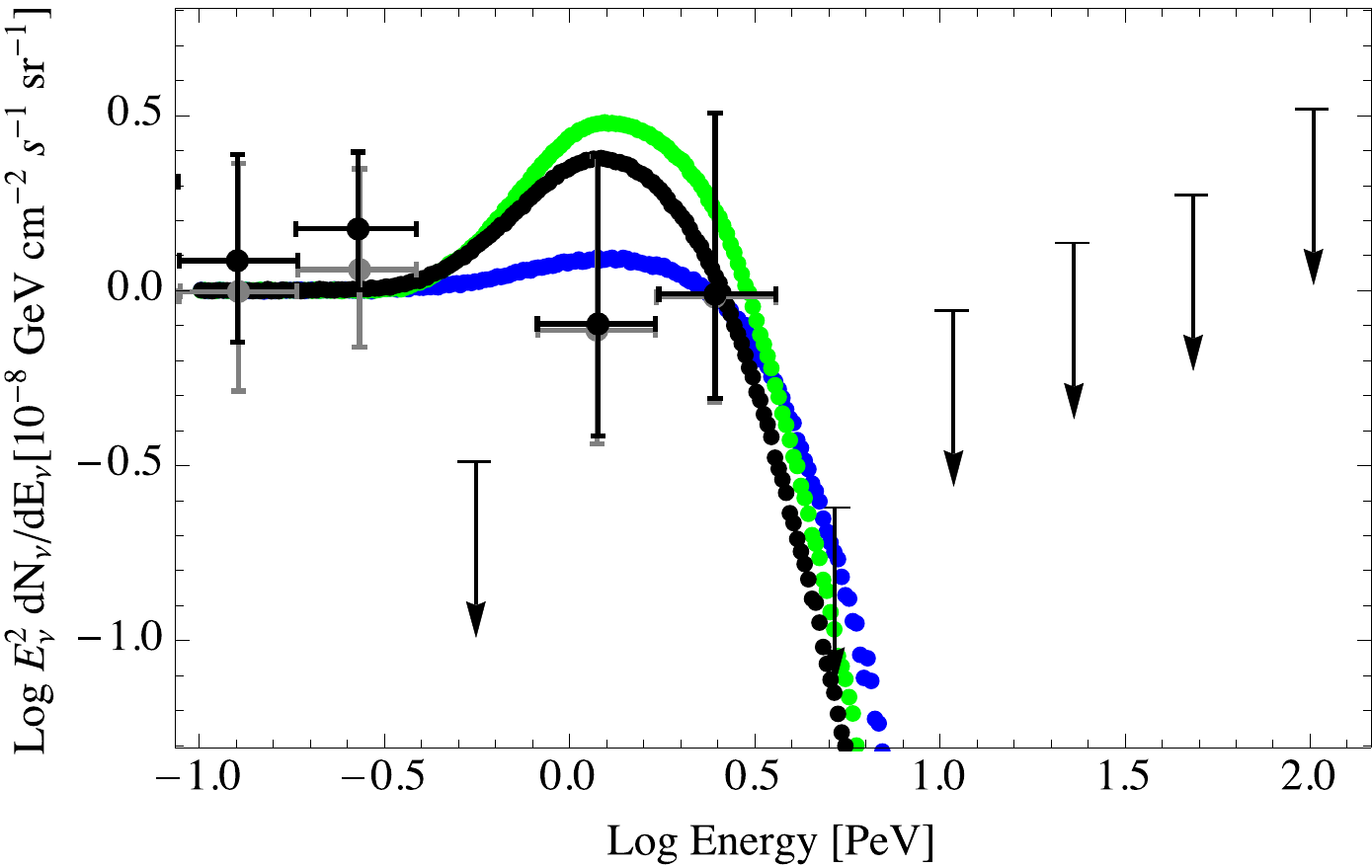}}
\caption{Propagated neutrino spectra including energy losses as described in the text\protect~\cite{st15}. Separately calculated n = 2 neutrino spectra with the VPE case shown in blue and the neutrino splitting case shown in green. The black spectrum takes account of all three processes
(redshifting, neutrino splitting, and VPE) occurring simultaneously. The rates for all cases are fixed by setting the rest frame threshold energy for VPE at 10 PeV. The neutrino spectra are normalized to the {\it IceCube} data both with (gray) and without (black) an estimated flux of prompt atmospheric neutrinos subtracted.~\cite{aa14}.}
\label{combined}
\end{figure}

\begin{figure}[!t]
\centerline{\includegraphics[width=4.3in]{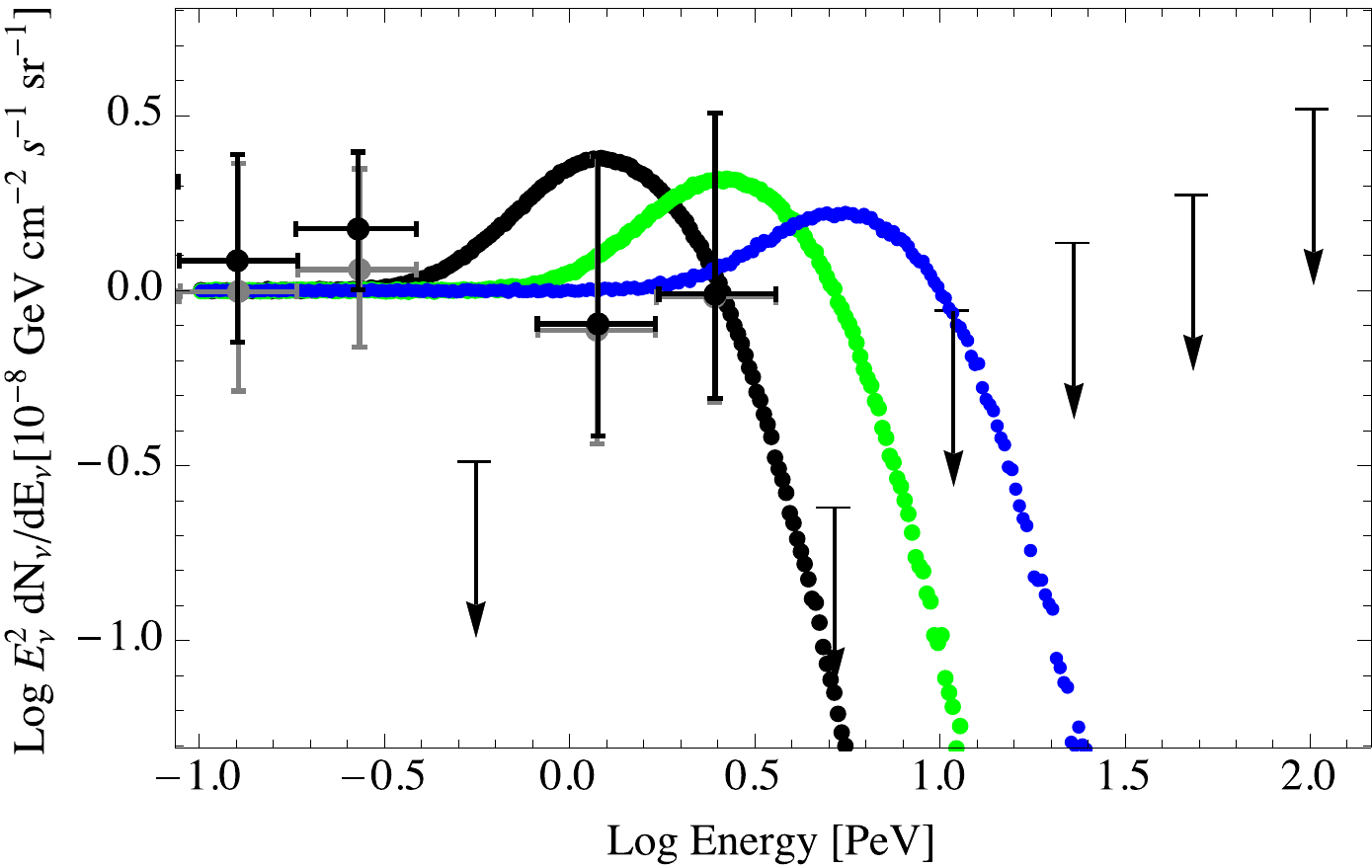}}
\caption{Calculated n = 2 spectra taking into account of all three processes
(redshifting, neutrino splitting, and VPE) occurring simultaneously for rest frame VPE threshold energies  of 10 PeV (black, as in Figure \ref{combined}), 20 PeV (green), and 40 PeV (blue). The {\it IceCube} data
are as in Figure~\ref{combined}~\cite{aa14}.}
\label{thresholdeffects}
\end{figure}

\subsection{[d] = 5 \cal{CPT} Violating Operator Dominance}

In the n = 1 case, the dominant $[d] = 5$ operator violates $\cal{CPT}$.
Thus, if the $\nu$ is superluminal, the $\bar{\nu}$ will be subluminal, and {\it vice
versa}. However, the {\it IceCube} detector cannot distinguish neutrinos from antineutrinos. 
The incoming $\nu (\bar{\nu}$) generates a shower in the detector, allowing
a measurement of its energy and direction. Even in cases where there is a muon
track, the charge of the muon is not determined. 

There would be an exception for electron antineutrinos at 6.3 PeV, 
given an expected enhancement in the event rate at the $W^{-}$ Glashow 
resonance since this resonance only occurs with $\bar{\nu_{e}}$. 
However, as we have discussed, no events have been detected
above 2 PeV. We note that $\nu - \bar{\nu}$ oscillation measurements would give
the strongest constraints on the difference in $\delta$'s between $\nu$'s and 
$\bar{\nu}$'s~\cite{ab15}. 

Since both VPE and neutrino splitting interactions generate
a particle-antiparticle lepton pair, one of the pair particles will be
superluminal ($\delta > 0$) whereas the other particle will be subluminal 
($\delta < 0$)~\cite{km13}. Thus, of the daughter particles,
one will be superluminal and interact, while the other will only redshift.
The overall result in the $[d] = 5$ case is that no clear spectral cutoff occurs~\cite{st15}.

\section{Summary of the results from LIV kinematic effects for superluminal neutrinos}
\label{sum}

In the SME EFT formalism~\cite{ko89,ck98,km13}, if the apparent cutoff above $\sim 2$ PeV in the neutrino spectrum shown in Figure \ref{thresholdeffects} is caused by LIV, this would result from an EFT with either a dominant $[d] = 4$ term with $\mathaccent'27 c^{(4)} = -\delta_{\nu e} = 5.2 \times 10^{-21}$, or by a dominant $[d] = 6$ term with $\mathaccent'27 c^{(6)} = -\kappa_2/M_{Pl}^2 \ge - 5.2 \times 10^{-35}$ GeV$^{-2}$~\cite{st15}. Such a cutoff would not occur if the dominant LIV term is a $\cal{CPT}$-violating the $[d] = 5$ operator. 

If the lack of neutrinos at the Glashow resonance is the result of LIV effects as shown in Figures
\ref{combined} and \ref{thresholdeffects}, this would imply that there will be no cosmogenic\cite{be69,st73} ultrahigh energy neutrinos. A less drastic effect in the cosmogenic neutrino spectrum can be caused by a violation of LIV in the hadronic sector at the level of $10^{-22}$\cite{sc11}.  

A cutoff can naturally occur if it is produced by a maximum acceleration energy in the sources. In that case, the parameters given above would be reduced to upper limits. However, the detection of a pronounced pileup just below the cutoff would be {\it prima facie} evidence of a $\cal{CPT}$-even LIV effect, possibly related to Planck-scale physics. In fact, $\cal{CPT}$-even LIV in the gravitational sector at energies below the Planck energy has been considered in the context of Ho\v{r}ava-Lifshitz gravity~\cite{ho09,po12}.

\section{Stable Pions from LIV}
\label{pi}

Almost all neutrinos are produced by pion decay. It has been suggested that if LIV effects can prevent the decay $\pi \rightarrow \mu + \nu$  of charged pions above a threshold energy, thereby eliminating higher energy neutrinos at the Glashow resonance energy and above~\cite{an14,to15}. In order for the pion to be stable above a critical energy $E_{c}$, we require that its effective mass as given by equation (\ref{effectivemass}) is less than the effective mass of the muon that it would decay to, i.e., $\tilde{m}_{\pi} < \tilde{m}_{\mu}$. This situation requires the condition that $\delta_{\pi} < \delta_{\mu}$~\cite{co99}. Then, neglecting the neutrino mass, this critical energy energy is given by
\begin{equation}
E_{c} = {{\sqrt{m_{\pi}^2 - m_{\mu}^2}}\over{\delta_{\mu} - \delta_{\pi}}}   
\end{equation}
\noindent noting that if, as before, we write $\delta_{\mu\pi} \equiv \delta_{\mu} - \delta_{\pi}$, then for small $\delta_{\mu\pi}$, it follows that $\sqrt{2\delta_{\mu\pi}} \simeq  \delta_{\mu\pi}$. In terms of SME formalism with Planck-mass suppressed terms, $\delta_{\mu\pi}$ is given by equation (\ref{d}) or equation (\ref{sub}). For example,
if we set $E_{c}$ = 6 PeV, in order to just avoid the Glashow resonance, we get the requirement,
$ \delta_{\mu\pi} \simeq 1.5 \times 10^{-8}$. 

If a lack of multi-PeV neutrinos is caused by this effect, there will be no pileup below the cutoff energy as opposed to the superluminal cases previously discussed. This would make the stable pion case difficult to distinguish from a natural cutoff caused by maximum cosmic ray acceleration energies in the neutrino sources.

\section{Observational tests with new neutrino telescopes}
\label{tel}

Future neutrino detectors are being planned or constructed: {\it IceCube-Gen 2}~\cite{aa14c}, the Askaryan effect detectors {\it ARA}~\cite{gu16} and {\it ARIANNA}~\cite{ba17}, and space-based telescopes such as {\it OWL}~\cite{st04}, {\it EUSO}~\cite{fe17}, and a more advanced OWL-type instrument called {\it POEMMA}. They will provide more sensitive tests of LIV. 

\section*{Acknowledgments} 
I would like to acknowledge my collaborators: Stefano Liberati, David Mattingly and Sean Scully. I thank Maria Gonzalez-Garcia and Michele Maltoni for helpful comments.

\end{document}